\documentclass[twocolumn,prl,showpacs,reprint]{revtex4-1}
\usepackage{amsmath}
\usepackage{graphicx}
\usepackage{color}

\begin{document}
\title{Short-time dynamics of monomers and dimers in quasi-two-dimensional colloidal mixtures}

\author{Erick Sarmiento-G\'omez$^{1}$}
\email{esarmiento@ifisica.uaslp.mx}
\author{Jos\'e Ram\'on Villanueva-Valencia$^{2}$}
\author{Salvador Herrera-Velarde$^{3}$}
\author{Jos\'e Arturo Ru\'iz-Santoyo$^{2}$}
\author{Jes\'us Santana-Solano$^{4}$}
\author{Jos\'e Luis Arauz-Lara$^{1}$}
\author{Ram\'{o}n Casta\~{n}eda-Priego$^{2}$}
\email{ramoncp@fisica.ugto.mx}

\affiliation{$^{1}$Instituto de F\'isica "Manuel Sandoval Vallarta", Universidad Aut\'onoma de San Luis Potos\'i, Alvaro Obreg\'on 64, 78000 San Luis Potos\'i, S.L.P., Mexico}

\affiliation{$^{2}$Divisi\'on de Ciencias e Ingenier\'ias, Campus Le\'on, Universidad de Guanajuato, Loma del Bosque 103, 37150, Le\'on, Mexico}

\affiliation{$^{3}$Subdirecci\'{o}n de Postgrado e Investigaci\'{o}n, Instituto Tecnol\'{o}gico Superior de Xalapa,
Secci\'on $5A$ Reserva Territorial \textit{s/n}, $91096$, Xalapa, Veracruz, Mexico}

\affiliation{$^{4}$Cinvestav Unidad Monterrey, Parque de Investigaci\'on e Innovaci\'on Tecnol\'ogica, Apodaca, Nuevo Le\'on 66629, Mexico}

\date{\today}
\begin{abstract}
We report on the short-time dynamics in colloidal mixtures made up of monomers and dimers highly confined between two glass-plates. At low concentrations, the experimental measurements of colloidal motion agree well with the solution of the Navier-Stokes equation at low Reynolds numbers, which takes into account the increase of the drag force on each particle due to wall-particle hydrodynamic forces. We find that the ratio of the short-time diffusion coefficients of the monomer and that of the center of mass of the dimer remains independent of both the total packing fraction and the dimer molar fraction up to concentrations near to the crystallization transition. The same physical scenario is observed for the ratio between the parallel and perpendicular components of the short-time diffusion coefficients of the dimer. This dynamical behavior is corroborated by means of Molecular Dynamics computer simulations that explicitly include the particle-particle hydrodynamic forces induced by the solvent. Thus, our results point out toward that the effects of the particle-particle hydrodynamic interactions on the diffusion coefficients are identical and, thus, factorable in both species. \end{abstract}

\maketitle

\textit{Introduction.} Many phenomena observed in colloidal dispersions resemble those of atomic systems. However, the colloidal dynamics exhibits special features due to the solvent-mediated forces typically known as hydrodynamic interactions (HI) \cite{Dhont1996}. Contrary to direct particle-particle interactions, HI can be tuned, but never completely screened or switched off. In a simple physical picture, HI can be understood as follows. The motion of a given colloidal particle induces a flow field in the solvent which is felt by the surrounding colloids. Thus, the motion of one colloidal particle causes a solvent-mediated force on the neighboring colloidal particles.

In contrast to the static counterpart, the colloidal dynamics is far from being completely understood. The reason is partially related with the fact that the colloidal dynamics extends over a wide range of temporal and length scales due to the enormous difference in size and mass between the colloids and the solvent molecules, giving rise to complex and long-ranged HI \cite{Dhont1996,Nagele1996}. The latter ones lead to non-trivial coupling among colloids that extends over many mean-interparticle distances \cite{Dhont1996,Nagele1996}. The understanding of HI is thus of relevance not only in physics, but also in several branches of science, such as biology, since phenomena like hydrodynamic synchronization in either biological systems (sperm, cilia, flagella) \cite{Kotar2010,Golestanian2011} or active fluids \cite{Damet2012}, and the dynamics of microswimmers \cite{Vifan2009,Yang2012} can only be explained in terms of hydrodynamic forces.

During the last few decades, the study of the hydrodynamic coupling between two or three spherical colloids \cite{Meiners1999,Herrera2013} or a spherical colloid near a wall \cite{Carbajal2007,Rice,Wajnryb} has been made possible thanks to the use of videomicroscopy techniques \cite{Xu2005} and the optical control of colloids through focused beams of light \cite{Meiners1999}. Besides, the contribution of HI on the colloidal dynamics at finite concentration has been mainly studied in model monodisperse suspensions composed of spherical colloids in the bulk \cite{Nagele1996} and confined between two plates or at the air-water interface \cite{Zahn1997,Santana2001,Santana2005,Dullens2015}. Nonetheless, a few years ago, the dynamical behavior of anisotropic colloids was experimentally studied at low particle concentrations \cite{Yodh,Weeks,Manoharan}. However, little is known about the effects of HI in multi-component systems made up of either spherical \cite{Valley2007} and non-spherical colloids. Recent contributions point mainly toward the study of the hydrodynamic coupling between the translational and rotational degrees of freedom of a family of non-spherical colloids \cite{Kraft2013} and the short-time diffusivity of dicolloidal particles \cite{Panczyk2014}. Nevertheless, nowadays, there is no a full understanding of the effects of the HI in multicomponent anisotropic colloidal systems.

Hence, in this Letter we report on the short-time dynamics in quasi-two-dimensional colloidal mixtures composed of monomers and dimers. Our results make evident that HI affect the short-time dynamics in such a way that the diffusion coefficient can be decomposed in two contributions, one including only confinement effects, in form of a diffusion coefficient at infinite dilution different from the bulk value, and a monotonically decreasing part that depends only on the volume fraction. Moreover, we give experimental and simulation evidence pointing out that the functional form of the latter contribution is identical and, thus, factorable, for both species at least for the range of concentrations studied here.

\begin{figure}[ht]
  \includegraphics[width=0.9\linewidth]{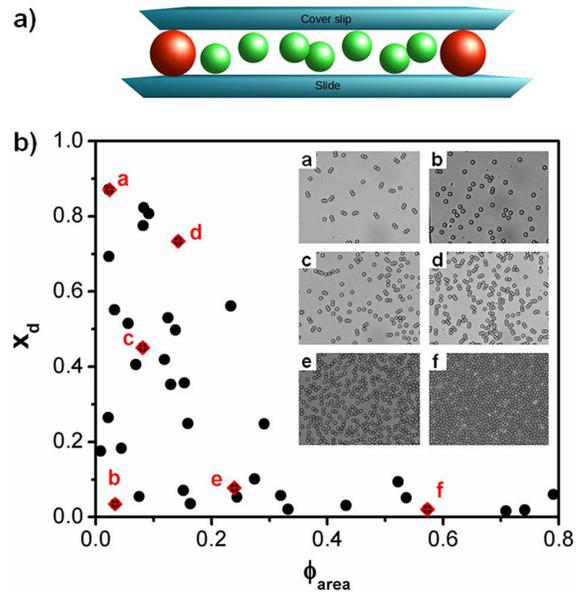}\\
  \caption{a) Cartoon of the colloidal mixture of monomers and dimers confined between two glass plates separated by spacers. b) Experimental (dots) and simulated (diamonds) points of the parameter space, $x_{d}$ vs $\phi$, considered in this work. Insets display snapshots of experimental equilibrium configurations.}\label{fig1}
\end{figure}

\textit{Experimental setup.} Colloidal dimers are prepared following the aggregation and fractionation procedure described in Ref. \cite{Arauz2008}. Polystyrene (spherical) particles in water solution of diameter $\sigma=$2 $\mu m$ (Duke Scientific) with negatively charged sulphate end groups on the surface are dialyzed against ultrapure water. Particle aggregation is promoted by the addition of $NaCl$ at a concentration of 500 $mM$ during six minutes. This process is quenched by dialysis of the colloidal suspension in clean water. To produce sinterization of particles in contact, a vial with the suspension is immersed in glycerol at a temperature close to the melting temperature of polystyrene (104 $^\circ$C) for 15 minutes. After cooling down at room temperature ($24\pm 0.1^\circ$ C), the different colloidal clusters are separated by centrifugation in a sugar density gradient tube; sugar is eliminated by centrifugation and redispersion in a aqueous solution of SDS at 1 $cmc$ to avoid further clusterization. The resulting dimers are now mixed with monomers and a small amount of larger particles of diameter 2.98 $\mu m$ that serve as spacers. The colloidal dispersion is confined between two clean glass plates (a slide and a cover slip). The system is sealed with epoxy resin and the mobile particles are allowed to equilibrate in the confined geometry. A schematic representation of the quasi-two-dimensional system is displayed in Fig. \ref{fig1}a.

In order to track the colloidal particles, the sample is observed from a top view, i.e., perpendicular to the glass walls, using an optical microscope with a 40 $\times$ objective and numerical aperture 0.6. The time evolution of the system is recorded using a standard video equipment \cite{Crocker}. Because the high confinement of the mixture, the motion of the particles perpendicular to the walls is highly restricted. Individual images extracted from the recorded video are analyzed to obtain both $x$ and $y$ coordinates of the center of the particles and from such data the trajectories are fully recovered. Dimers are identified by tracking the average distance between adjacent particles as a function of time. We found an average distance between center of particles forming dimers to be $0.9$ $\pm 0.04$ $\sigma$, as consequence of the sinterization procedure. From the trajectories, both translational and rotational components of the mean-square displacements (MSDs) are obtained. In the case of the dimers, the translational MSD can also be decomposed in the directions parallel and perpendicular to the long axis of symmetry.

The previous protocol is carried out for several dimer molar fractions $x_{d}$ and total packing fractions $\phi$ in order to explore a wide number of points in the parameters space, as displayed in Fig. \ref{fig1}b. For the comparison of the experimental measurements with computer simulations, we have selected six representative points of the parameters space (diamonds in Fig. \ref{fig1}b labelled as a,b,c,d,e,f).
%
%
%

\textit{Diffusion at low concentrations: wall-particle hydrodynamic forces.} It is well-known that HI and the Brownian motion of colloidal particles are dramatically affected by the presence of walls \cite{Carbajal2007,DeCorato2015}. In particular, several experiments have been carried out to understand the single-particle diffusion near to a flat wall \cite{Rice2000,Happel1983,Brenner1961}. Furthermore, Rice and co-workers derived a robust approximation to determine the decrease of the diffusion coefficient parallel to the walls, which was corroborated in experiments of an isolated colloidal sphere confined between flat plates \cite{Rice2000}. For the conditions of our experiments, such approximation predicts a reduction of $45\%$ in the short-time diffusion coefficient, $D^{s}_{0}$, of a single sphere at the bulk, i.e., the so-called Stokes-Einstein diffusion coefficient, $D^{s}_{0}=k_{B}T/3\pi\eta\sigma$, where $k_{B}$ is the Boltzmann constant, $T$ the absolute temperature and $\eta$ the viscosity. This prediction is close to the experimental value measured for the monomers at very diluted conditions (state b, Fig. \ref{fig1}b), whose reduction is about $35\%$ the value of $D^{s}_{0}$.

As mentioned previously, the dynamics of non-spherical colloids is nowadays a challenging topic of paramount importance within the context of condensed matter physics. From theoretical point of view, recent advances have been achieved by solving the Navier-Stokes equations in the limit of low Reynolds number, $Re \ll 1$, for anisotropic particles \cite{Zabarankin,DeCorato2015}. In particular, Zabarankin \cite{Zabarankin} calculated the drag force and the resisting torque experienced by a dicolloid-like particle at the bulk. Then, using Zabarankin's approximation together with a similar approach as the one proposed by Rice \cite{Rice2000}, a reduction of $45\%$ in the short-time diffusion coefficient of the center of mass of a dimer is estimated. One can compare this result with the experimental measurements, also in the limit of low concentrations (state a, Fig. \ref{fig1}b). The reduction turns out to be only $30\%$ the bulk value. The difference between theory and experiments is larger than in the case of the monomers. This might be linked to two sources that are not taken into account in the theoretical description: i) the dimer is not located at the center of the glass plates and ii) it is able to rotate out the parallel plane to the walls. Nonetheless, Happel and Brenner \cite{Happel1983} have studied the dynamical behavior of a single spheroid located in the middle of two parallel walls. Then, by mapping the characteristic dimensions of the dimer used in the experiments directly to the spheroid, it is possible to find a decrease in the transverse diffusivity of $33\%$ (theory) and $39\%$ (experiments), which is in better agreement with our estimations and measurements.

\textit{Computer simulations and temporal re-scaling of the short-time dynamics.}
To confirm our experimental results at higher concentrations, we have carried out Molecular Dynamics (MD) simulations of the whole colloidal
dispersion taking into account explicitly the solvent molecules; this avoids the use of an approximation for the colloid-colloid
hydrodynamic interactions. The simulation results have been produced with the open source
MD package ESPResSo \cite{imbach06a,arnold13a}. In the simulations, we have considered a $2D$ system made up of
monomers and dimers in a bath of smaller particles representing the solvent molecules. All
particles interact through a pseudo-hard-sphere model recently proposed by Jover et al. \cite{Jackson}; solvent particles
have a diameter ten times smaller than the colloids size and occupy a surface fraction that allows us to match
the experimental conditions, however, in a few cases, we have checked that the same results are obtained when the solvent molecules are twenty times smaller (data not shown). Dimers are formed
using the rigid bond feature of ESPResSo. At the beginning of the simulation, all particles are randomly distributed in the simulation box
and during the runs the center of mass of each particle is constrained to move on the $xy-$plane, i.e., motion along the $z-$direction
is not allowed. Simulations are performed in the $NVT$ ensemble using the Langevin thermostat implemented in ESPResSo. The total number of
monomers and dimers used in the simulation matches the number of colloids observed in the view field of the experiment,
ranging from $N = 50$ up to $N = 950$ colloids. Since the computer simulation with the explicit inclusion of solvent molecules is a time-consuming
task \cite{Torres2015}, we only simulate those state points indicated by diamonds in Fig. \ref{fig1}b.

We should point out that the dynamics provided by the MD simulation does not lie in the same temporal regime of the experiments. Nevertheless, it is possible to re-scale the MD time to directly compare with the experiments \cite{Lopez2013,Perez2011}. The re-scaling consists in replacing the MD time as follows: $t \rightarrow \sqrt{k_{B}T/M}\frac{\sigma}{D^{0}_{s}}t$, where $M$ is the mass of a monomer. Hence, we present all the simulation results with such temporal re-scaling.
\begin{figure}[ht]
  \includegraphics[width=0.9\linewidth]{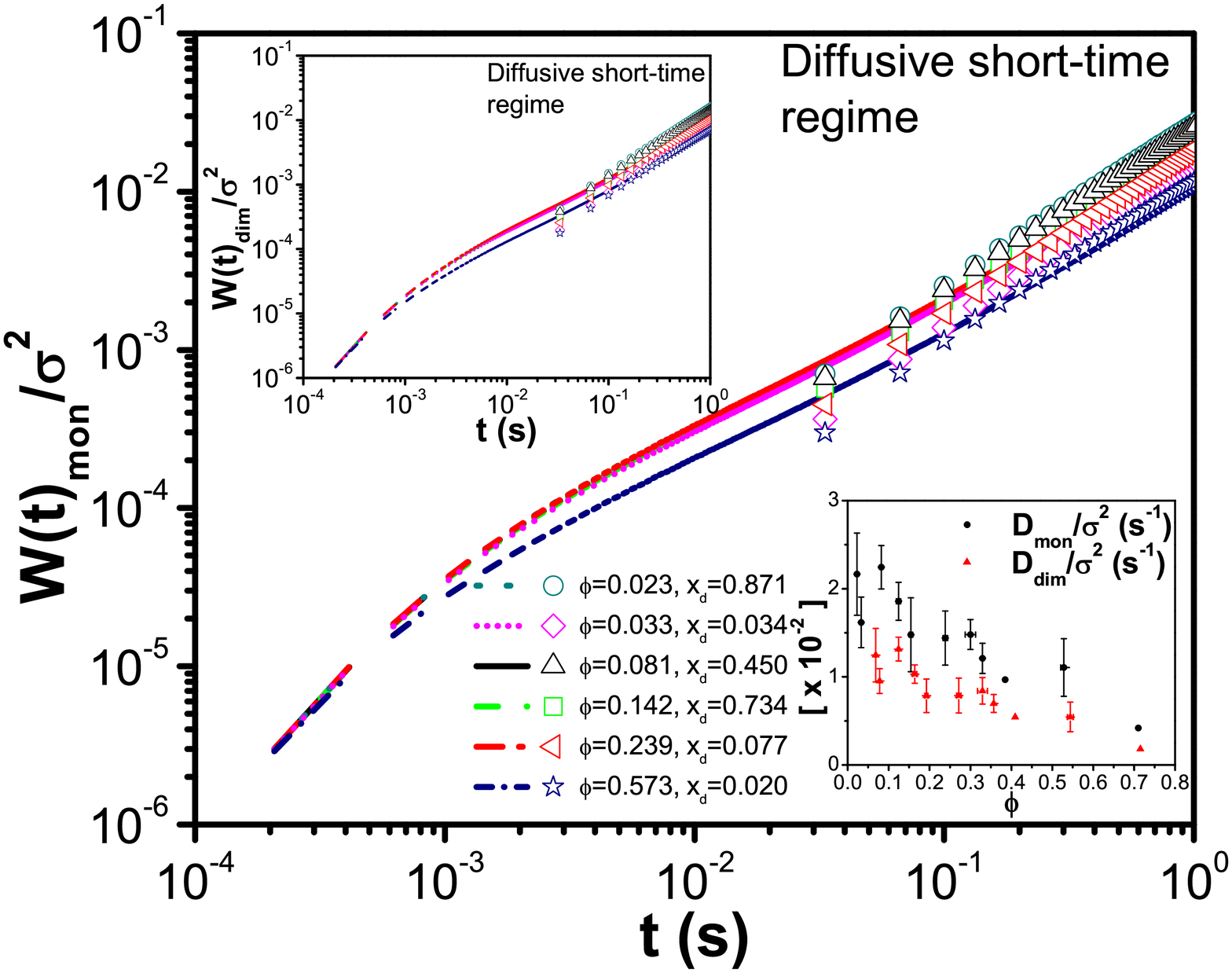}\\
  \caption{Mean-square displacement, $W(t)$, of the center of mass of a monomer (main body) and the center of mass of a dimer (inset-left). Inset-right displays the short-time diffusion coefficients, $D_{s}$, of both species as a function of the total packing fraction obtained from the expression: $W(t)=4 D_{s}t$. Symbols are experimental data and lines simulation results for those state points indicated by diamonds in Fig. \ref{fig1}b.}\label{fig2}
\end{figure}

\textit{Diffusion at finite concentrations: colloid-colloid hydrodynamic forces.} The mean-square displacements, $W(t)\equiv \left<(\vec{r}(t)-\vec{r}(0))^2\right>$, for the monomer (main body) and the center of mass of the dimer (inset-left) for those state points indicated in Fig. \ref{fig1}b are displayed in Fig. \ref{fig2}. As it is seen from the figure, the MD results clearly display the ballistic ($W(t) \propto t^{2}$) and diffusive ($W(t) \propto t$) regimes characteristic of the underlying Newtonian dynamics. Overall, one can observe a very good agreement between MD simulations and experiments. This confirms that the simulation model captures correctly the colloid-colloid hydrodynamic effects and that the temporal re-scaling of the MD simulations is appropriated. Besides, we should point out that this level of agreement makes evident that the hydrodynamic interactions are indispensable to explain the observed diffusive behavior, otherwise the colloidal dynamics is not correctly captured (data not shown). Furthermore, as expected, one can observe that the monomers diffuse faster than the dimers. Inset-right shows that the short-time diffusion coefficients of both species decrease monotonically with the concentration following the same trend. This suggests that the effects of the HI on the short-time diffusion coefficient can be decomposed in two contributions: one that does not depend on the volume fraction, but only on the confinement (as explained above), and a contribution that is a function of the concentration. Inset-right also indicates that the concentration-dependence of the short-time diffusion coefficients has a similar functional form in both species. We discuss an important consequence of this observation further below.

\begin{figure}[ht]
  \includegraphics[width=0.9\linewidth]{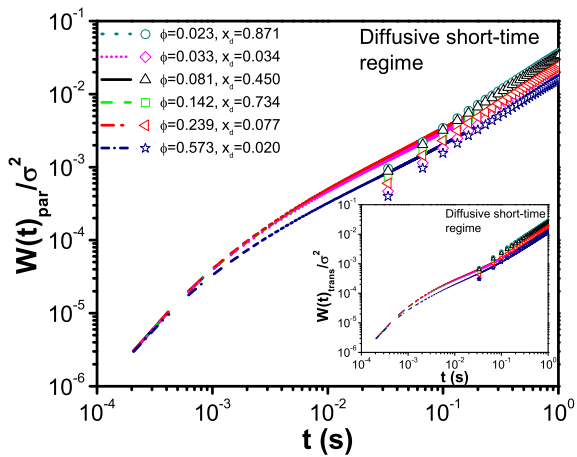}\\
  \caption{Mean-square displacement, $W(t)$, for the parallel (main body) and perpendicular (inset) displacements, respect to the main axis, of a dimer. Symbols are experimental data and lines simulation results for those state points indicated by diamonds in Fig. \ref{fig1}b.}\label{fig3}
\end{figure}

As mentioned above, the MSD of the dimers can also be decomposed in two contributions: parallel and perpendicular to the main axis of symmetry. In Fig. \ref{fig3} such contributions are explicitly displayed for the state points indicated in Fig. \ref{fig1}b. There, one can appreciate that the short-time diffusion coefficients, parallel (main body) and perpendicular (inset), decrease when the concentration increases. One can also notice that the diffusion along the parallel direction becomes faster than the perpendicular one. This can be easily understood if one considers that the cross section perpendicular to the axis of symmetry of the dimer is smaller than the parallel one. This means that less solvent molecules collide frontally with the dimer, then reducing drastically the friction in the parallel direction. The opposite scenario occurs in the perpendicular direction. Nonetheless, at sufficiently long-times, i.e., when the dimer has explored most of its phase space, both MSDs reach the same value (data not shown). Again, within the time window shown here, the computer simulations are able of capturing the whole diffusive mechanisms.

\begin{figure}[ht]
  \includegraphics[width=0.9\linewidth]{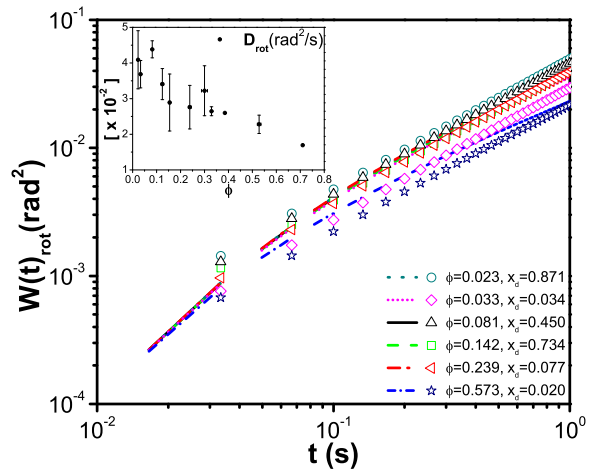}\\
  \caption{Mean-square angular displacement of a dimer, $W(t)_{rot}$. Symbols are experimental data and lines simulation results for those state points indicated by diamonds in Fig. \ref{fig1}b. Inset shows the density-dependence of the short-time  angular diffusion coefficient.}\label{fig4}
\end{figure}

To fully characterize the dynamics of the dimers, we have also measured the mean-square angular displacement (MSAD), which is explicitly shown in Fig. \ref{fig4}. From this figure, one can observe that the computer simulations predict a MSAD that displays a linear-time dependence at short-times ($W(t)_{rot} \propto t$). At this regime, the simulations capture the characteristic rotational diffusion of an ellipsoidal Brownian particle confined between two flat walls \cite{Neild2010} and the tendency seen in the translational dynamics discussed above: the rotational diffusion coefficient decreases with the total packing fraction (see inset). In this case, the deviations between simulations and experiments are more notorious; this difference might be related with the fact that the temporal re-scaling used to compare the simulation data with the experimental results is not completely correct to account for the rotational dynamics.

\begin{figure}[ht]
  \includegraphics[width=0.8\linewidth]{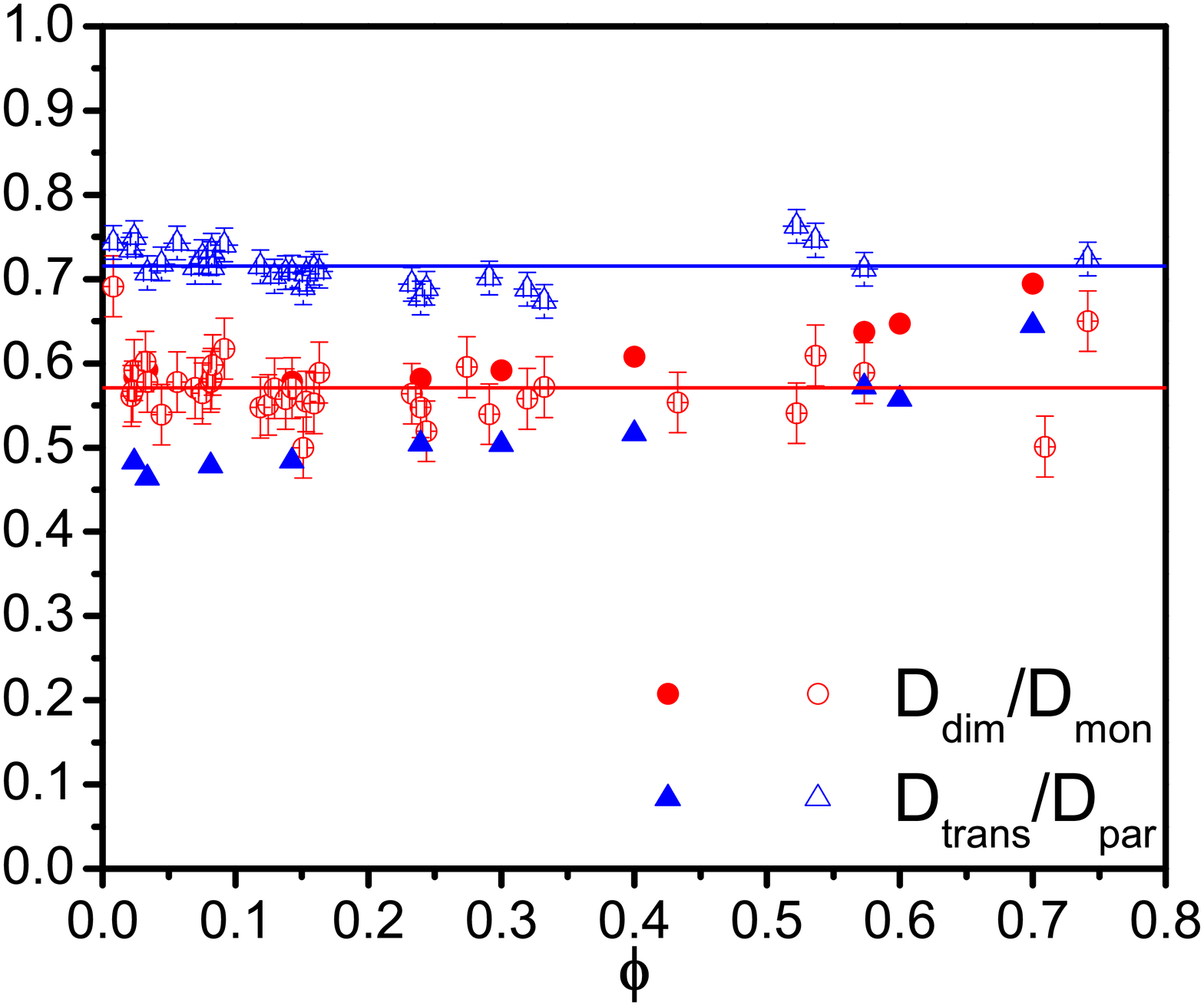}\\
  \caption{Experimental (open symbols) and simulation (closed symbols) results for the ratio of the short-time diffusion coefficients of the center of mass of the dimer and monomers, and the ratio of the parallel and transversal short-time diffusion coefficients of a dimer as a function of the total packing fraction. Solid lines just guide the eye.}\label{fig5}
\end{figure}

\textit{Discussion: decomposition and factorization.} So far, we have characterized the dynamics of both species, i.e., monomers and dimers, in a wide interval of concentrations (see Fig. \ref{fig1}b). As explained above, the wall-particle hydrodynamic forces are responsible for the increase of the drag force parallel to the walls. This mechanism clearly affects the particle diffusion in the parallel plane, but it is an effect that enters in the same way on each particle regardless the type of species, whereas the diffusion in the perpendicular direction is suppressed due to the high degree of confinement of the particles. Thus, the phenomenology observed in the MSDs should be associated to a competition between the direct forces among colloids and the colloid-colloid hydrodynamic forces. Naively, one might think that the latter ones do not contribute dramatically at low and high particle concentrations, but should be important at intermediate concentrations. Additionally, due to the torques that the dimers experience, one can also think that the translational-rotational hydrodynamic coupling may affect drastically their dynamical behavior \cite{Neild2010}. Therefore, in order to quantify the difference in the diffusivity of monomers and dimers, we consider the ratio of the short-time diffusion coefficients of the center of mass of the dimer and the monomer. This ratio is explicitly shown in Fig. \ref{fig5} as a function of the total packing fraction. Interestingly, such ratio is practically independent of both the total packing fraction and the molar fraction of dimers and takes the value $0.57\pm 0.03$. This means that hydrodynamic interactions play the following crucial role in the colloidal dynamics of the mixture: their effects on the diffusion coefficients are identical in both species and can be decomposed in a short time coefficient including only a confinement contribution, and a monotonically decreasing concentration-dependent part; a similar conclusion has been reached in experiments of near-wall dynamics of concentrated hard-spheres suspensions \cite{Liu2015}. In fact, computer simulations predict the same physical scenario. A similar behavior is also seen for the ratio of the transversal and parallel short-time diffusion coefficients of the dimer, finding an experimental value of $0.71 \pm 0.02$, although in this case the simulations ($0.52 \pm 0.05$) underestimate the value measured in the experiments. This difference might be related to the number of approximations included in the simulation model that does not take into account several features of the physical system, for example, dimers can slightly rotate along the perpendicular plane of confinement (producing an interaction with the walls) and the proper estimation of the solvent viscosity.

\textit{Conclusions.} We have demonstrated that the ratio of the short-time self-diffusion coefficients in a colloidal mixture composed of monomers and dimers is independent of the particle concentration up to a surface fraction of $0.70$. Our findings imply that hydrodynamic interactions are equal for both species and, in consequence, they become factorable. Thus, by characterizing the short-time self-diffusion of one of the species, the other one can be straightforward determined. However, further experiments and computer simulations are needed to corroborate this scenario in the case of mixtures with either higher size ratios or different shapes. Our findings are thus crucial for the understanding of the colloidal dynamics and for the development of theoretical frameworks that take into account hydrodynamic interactions in anisotropic mixtures at finite concentration.

\begin{acknowledgments}
Authors acknowledge partial financial support from Conacyt (Grant Nos. 237425 and 182132). R. C.-P. also acknowledges the financial support provided by the Marcos Moshinsky fellowship 2013-2014 and the University of Guanajuato (Convocatoria Institucional para Fortalecer la Excelencia Acad\'emica 2015).
\end{acknowledgments}


\begin{thebibliography}{99}
\bibitem{Dhont1996} J. K. G. Dhont, An Introduction to Dynamics of Colloids, Elsevier, 1996.
\bibitem{Nagele1996} G. N\"agele, Phys. Rep. 272, 215 (1996).
\bibitem{Kotar2010} J. Kotar, M. Leoni, B. Bassetti, M. C. Lagomarsino, P. Cicuta, Proc. Natl. Acad. Sci. 107, 7669 (2010).
\bibitem{Golestanian2011} R. Golestanian, J. M. Yeomans, and N. Uchida, Soft Matter 7, 3074 (2011).
\bibitem{Damet2012} L. Damet, G. M. Cicuta, J. Kotar, M. C. Lagomarsino and P. Cicuta, Soft Matter 8, 8672 (2012).
\bibitem{Vifan2009} A. Vifan and H. Stark, Phys. Rev. Lett. 103, 199801 (2009).
\bibitem{Yang2012} W. Yang, V. R. Misko, K. Nelissen, M. Kong and F. M. Peeters, Soft Matter 8, 5175 (2012).
\bibitem{Meiners1999} J. C. Meiners and S. R. Quake, Phys. Rev. Lett. 82, 2211 (1999).
\bibitem{Herrera2013} S. Herrera-Velarde E. C. Eu\'an-D\'iaz, F. C\'ordoba-Vald\'es and R. Casta\~neda-Priego, J. Phys.: Cond. Matt. 25, 325102 (2013).
\bibitem{Rice} B. Lin, J. Yu, S. A. Rice. Phys. Rev. E 62, 3909 (2000).
\bibitem{Wajnryb} S. Bhattacharya, J. Blawzdziewicz, and E. Wajnryb, J. Fluid Mech. 541, 263 (2005).
\bibitem{Carbajal2007} M. Carbajal-Tinoco, R. L\'opez-Fern\'andez and J. L. Arauz-Lara, Phys. Rev. Lett. 99, 138303 (2007).
\bibitem{Xu2005} X. Xu et al., Phys. Rev. Lett. 95, 158301 (2005).
\bibitem{Zahn1997} K. Zahn, J. M. M\'endez-Alcaraz, and G. Maret, Phys. Rev. Lett. 79, 175 (1997).
\bibitem{Santana2001} J. Santana-Solano and J. L. Arauz-Lara, Phys. Rev. Lett. 87, 038302 (2001).
\bibitem{Santana2005} J. Santana-Solano, A. Ram\'irez-Saito and J. L. Arauz-Lara, Phys. Rev. Lett. 95, 198301 (2005).
\bibitem{Dullens2015} A. L. Thorneywork, R. E. Rozas, R. P. A. Dullens and J. Horbach, in press Phys. Rev. Lett. (2015).
\bibitem{Yodh} Y. Han, A. M. Alsayed, M. Nobili, J. Zhang, T. C. Lubensky, A. G. Yodh, Science 314, 626 (2006).
\bibitem{Weeks} K. V. Edmond, M. T. Elsesser, G. L. Hunter, D. J. Pine, E. R. Weeks, Proc. Natl. Acad. Sci. 109, 44, 17892 (2012).
\bibitem{Manoharan} A. Wang, T. G. Dimiduk, J. Fung, S. Razavi, I. Kretzschmar, V. N. Manoharan, J. Quan. Spec. Rad. Trans. 146, 499 (2014).
\bibitem{Valley2007} D. T. Valley, S. Rice, B. Cui, H. Diamant and B. Lin., J. Chem. Phys. 126, 134908 (2007).
\bibitem{Kraft2013} D. J. Kraft, R. Wittkowski, B. ten Hagen, K. V. Edmond, D. Pine and H. L\"owen, Phys. Rev. E 88, 050301(R) (2013).
\bibitem{Panczyk2014} M. M. Panczyk, N. J. Wagner and E. M. Furst, Phys. Rev. E 89, 062311 (2014).
\bibitem{Arauz2008} A. Garcia-Castillo, and J. L. Arauz-Lara: Phys. Rev. E 78, 020401(R) (2008).
\bibitem{Crocker} J. C. Crocker and D. G. Grier, J. Coll. Int. Sci., 179, 298, (1996).
\bibitem{DeCorato2015} M. De Corato, F. Greco, G. D'Avino and P. L. Maffettone, J. Chem. Phys. 142, 194901 (2015).
\bibitem{Rice2000} B. Lin, J. Yu, S. A. Rice: Coll. Surf. A 174, 121 (2000).
\bibitem{Brenner1961} H. Brenner, Chem. Eng. Sci. 16, 242 (1961).

\bibitem{Happel1983} J. Happel and H. Brenner, Low Reynolds Number Hydrodynamics, Kluwer, Dordrecht, 1983.

\bibitem{Zabarankin} M. Zabarankin, Proc. R. Soc. A 463, 2329 (2007).

\bibitem{imbach06a} H. J. Limbach, A. Arnold, B. A. Mann and C. Holm, Comp. Phys. Comm. 174, 704 (2006).

\bibitem{arnold13a} A. Arnold et al., ESPResSo 3.1--- Molecular Dynamics Software for Coarse-Grained
 Models, Meshfree Methods for Partial Differential Equations VI, 89, 1-23 (2013).


\bibitem{Jackson} J. Jover, A. J. Haslam, A. Galindo, G. Jackson and E. A. M\"uller, J. Chem. Phys. 137, 144505, (2012).
\bibitem{Torres2015} A. Torres-Carbajal, S. Herrera-Velarde and R. Casta\~neda-Priego, Phys. Chem. Chem. Phys 17, 19557 (2015)
\bibitem{Lopez2013} L. L\'opez-Flores, H. Ru\'iz-Estrada, M. Ch\'avez-P\'aez and M. Medina-Noyola, Phys. Rev. E 88, 042301 (2013).
\bibitem{Perez2011} G. P\'erez-Angel et al., Phys. Rev. E 83, 060501(R) (2011).
\bibitem{Neild2010}  A. Neild et al., Phys. Rev. E 82, 041126 (2010).
\bibitem{Liu2015}  Y. Liu et al., Soft Matter 11, 7316 (2015).



\end{thebibliography}
\end{document}